\documentstyle[12pt,epsf]{article}

\begin{document}

\begin{titlepage}

\begin{flushright}
HUPD-9403\\
March, 1994\\
\end{flushright}

\vspace*{1.5cm}
\begin{center}

{\large
   Computation of the Heavy-Light Decay Constant using
   Non-relativistic Lattice QCD \\[1cm]
}

   Shoji Hashimoto \\[5mm]

{\sl
   Department of Physics, Hiroshima University\\
   Higashi-Hiroshima, Hiroshima 724\\
   Japan\\
}
\end{center}

\vspace*{2cm}
\begin{center}
{\bf
   abstract
}
\end{center}

We report results on a lattice calculation of the heavy-light meson
decay constant employing the non-relativistic QCD approach
for heavy quark and Wilson action for light quark.
Simulations are carried out at $\beta=6.0$ on a $16^3\times 48$ lattice.
Signal to noise ratio for the ground state is
significantly improved compared to simulations in the static approximation,
enabling us to extract the decay constant reliably.
We compute the heavy-light decay constant for
several values of heavy quark mass
and estimate the magnitude of the deviation
from the heavy mass scaling law $f_{P} \sqrt{m_{P}} = const$.
For the $B$ meson we find $f_{B} = 171\pm 22^{+19}_{-45}$ MeV,
while an extrapolation to the static limit yields
$f_{B}^{static}$ = $297\pm 36^{+15}_{-30}$ MeV.

\end{titlepage}

\setlength{\baselineskip}{18pt}

\section{Introduction}

Computation of electro-weak transition matrix elements of hadrons 
is one of the most important goals of numerical simulations of
Lattice QCD.
In particular the decay constant of heavy-light mesons
which consist of one heavy quark ($Q$) and one light anti-quark ($q$)
has great phenomenological and theoretical interest in that finding its  
value represents a crucial step to extract the Cabibbo-Kobayashi-Maskawa 
matrix elements from experiment.

For heavy-light mesons such as  B and D  the method of heavy quark
effective theory \cite{HQET_review} is applicable.
In the static limit in which the heavy quark field is replaced
by its infinite mass limit,
implementation of the heavy quark effective theory on
the lattice is straightforward
and was first proposed by Eichten\cite{Eichten_1}
and by Lepage and Thacker\cite{Lepage_Thacker}.

The initial attempt to simulate heavy-light systems 
in the static limit was made by Boucaud et al.\cite{Ph.Boucaud_etal_1}.  
They found that the signal of the heavy-light meson
correlation function is extremely noisy,
which made it impossible to identify a plateau
in the effective mass corresponding to the ground state meson.
In order to avoid this problem several 
groups\cite{Alexandrou,Allton_1,Eichten_2,Bernard_1}
applied the ``smeared'' source technique,
with which the signal for the ground state is expected to be enhanced.
These groups reported  large values of $f_{B}$
in a range $f_{B}^{static} =$ 300-500 MeV.
However, it was pointed out by two groups\cite{Hashimoto_Saeki,Bernard_2}
that the extracted values of $f_{B}$ strongly correlate
with the size of regions
over which the source of the heavy-light meson is smeared.
This systematic effect casts doubt on the reliability of the results
from the static approximation.
It is most likely that one has failed to identify the correct
ground state when one used arbitrarily chosen smeared sources.

The reason why large statistical fluctuations
occur when one applies the static approximation has been explained
by Lepage\cite{Lepage} with a very simple way of
estimating the statistical noise
of the hadron correlation functions.
He pointed out at the same time that
the use of heavy quark effective theory including the kinetic term
should decrease the statistical noise in correlation functions.  
Taking into account the kinetic and other $1/m_Q<$ terms
where $m_Q$ is the heavy quark mass is also necessary
to estimate the size of $1/m_Q$ corrections in physical quantities.
This in fact is very important  to extract physical
predictions for the $B$ meson decay constant. 

The heavy quark effective theory including $1/m_Q$ terms is called
non-relativistic QCD.
In this paper we report results of a lattice calculation of
the heavy-light decay constant using non-relativistic QCD for heavy quark
and Wilson action for light quark.
Our simulations are carried out in the quenched
approximation on a $16^3\times 48$ lattice at $\beta=6/g^{2}=6.0$.
We examine to what extent the inclusion of $1/m_Q$ terms improve signals for
the ground state over those in the static limit,
and check the independence of results for
the decay constant on the smearing size.
Employing a set of heavy quark masses in the range $m_Qa=10-2.5$,
we examine the magnitude of deviation from the heavy quark scaling law
$f_P\sqrt{m_P}=$constant.
Combining the results of this analysis with
an estimate of the renormalization factors including tadpole
improvement\cite{Lepage_etal_1},
we extract physical prediction for the $B$ meson decay constant.  

This paper is organized as follows.
In sect.~2 we briefly review the formulation of non-relativistic lattice QCD.
In sect.~3 we describe details of our simulations
and present our results for the signal to noise ratio
and the heavy-light decay constant.
In sect.~4 we discuss perturbative determination
of renormalization factors,
and present our results for the physical value 
of the $B$ meson decay constant.
Our conclusions are given in Sec.~5.

\section{Non-relativistic Lattice QCD}

The general form of the Lagrangian of heavy quark effective theory is given 
as a power series in the inverse heavy quark mass $m_{Q}^{-1}$:
\begin{equation}
{\cal L} = {\cal L}_{0} + {\cal L}_{1} + \cdots
\end{equation}
The first two terms are given by 
\begin{eqnarray}
{\cal L}_{0} & = & Q^{\dagger} i D_{0} Q \\
{\cal L}_{1} & = & Q^{\dagger} [
                   \frac{\mbox{\boldmath $D$}^{2}}{2 m_{Q}}
                   + c \frac{g}{2 m_{Q}}
                       \mbox{\boldmath $\sigma$} \cdot \mbox{\boldmath $B$}
                 ] Q
\end{eqnarray}
where $Q(x)$ is a two component heavy quark field, 
$D_{\mu}$ the color SU(3) covariant derivative, and 
{\boldmath $B$} denotes the chromomagnetic field.
The leading order term ${\cal L}_{0}$ represents the infinite mass
or the static limit.
The next to leading order term ${\cal L}_{1}$ consists of the non-relativistic 
kinetic energy and the Pauli
spin-magnetic interaction for heavy quark. For studies of heavy-heavy systems
($\Psi$'s and $\Upsilon$'s) a power counting rule in terms of the heavy quark
velocity instead of the inverse heavy quark mass is
applicable\cite{Lepage_etal_1}.  Since  the spin-magnetic interaction term gives
higher order effects in the heavy quark velocity,
one can omit this term.
We are, however, interested in the heavy-light system
for which the heavy quark velocity is not a good expansion parameter.  
One therefore must properly include the spin-magnetic interaction term,
and we treat the heavy quark effective lagrangian including
all $1/m_{Q}$ terms.

The lattice action we use for simulations is given by
\begin{equation}
S_{Q} = \sum_{\mbox{\boldmath $x$} t}
        Q^{\dagger}(x)
        [ \Delta_{4}
        + H^{(n)}
        +c \frac{g}{2 m_{Q}}
                       \mbox{\boldmath $\sigma$} \cdot \mbox{\boldmath $B$}
        ] Q(x)
\label{action}
\end{equation}
where
\begin{eqnarray}
H^{(n)} & = & 1 - ( 1 - \frac{H}{n} )^{n},\\
H & = & -\sum_{j=1}^{3} \frac{\Delta_{-j} \Delta_{j}}{2 m_{Q}},
\end{eqnarray}
The lattice spacing is denoted by $a$ and
the gauge covariant difference operator
$\Delta_{\mu}$ is defined by 
\begin{eqnarray}
\Delta_{\mu} Q(x) & = & U_{\mu}(x) Q(x+\hat{\mu}) - Q(x) \nonumber \\
\Delta_{-\mu} Q(x) & = & Q(x) - U_{\mu}^{\dagger}(x-\hat{\mu}) Q(x-\hat{\mu})
\end{eqnarray}
where $U_{\mu}(x)$ is the lattice gauge link variable.
The action with $n\geq 2$ represents a modification in order to stabilize
high frequency modes for $m_{Q} < 3/n$\cite{Thacker_Lepage}.
The spin-(chromo)magnetic interaction term is included for keeping
consistency of the $1/m_{Q}$ expansion
even though it is not expected to give
much effect on the pseudo-scalar decay constant.  The usual clover-leaf
definition is employed for the chromo-magnetic field operator $\mbox{\boldmath
$B$}$. For the coupling of the spin-magnetic interaction term we use the tree
level value $c$=1.

The 4-component Dirac field of the heavy quark is expressed
in terms of the 2-component field $Q(x)$ as
\begin{equation}
\Psi(x) = \left(
          \begin{array}{c}
          Q(x) \\
          \frac{-i}{2 m_{Q}}
          \mbox{\boldmath $\sigma$} \cdot \mbox{\boldmath $B$}
          Q(x)
          \end{array}
          \right)
          +O(1/m_Q^2).
\end{equation}
We omit the lower components for heavy quark for simplicity and define bilinear
operators composed of the heavy and light quarks as
\begin{equation}
{\cal O}_{\Gamma}(x) = (
                       \begin{array}{cc}
                       Q^{\dagger}(x), & 0
                       \end{array}
                       ) \Gamma q(x).
\end{equation} 
For instance the lattice axial vector current is given by
\begin{equation}
A_{\mu}(x) = (
              \begin{array}{cc}
              Q^{\dagger}(x), & 0
              \end{array}
             ) \gamma_{\mu} \gamma_{5} q(x).
\label{heavy-light_current}
\end{equation}

The heavy quark propagator is obtained solving the evolution equation
\begin{equation}
G(\mbox{\boldmath $x$},t+1) = U^{\dagger}_{x,t,4}
            \left[\left( 1- \frac{1}{n} H \right)^{n}
            G(\mbox{\boldmath $x$},t)+\Phi(\mbox{\boldmath $x$})\delta_{t,0}
          \right]
\end{equation}
where $\Phi(\mbox{\boldmath $x$})$ is the source function of the heavy quark.
Computation of this deterministic evolution equation is much faster than
solving the inverse of the Wilson quark operator.

\section{Simulation}

\subsection{Simulation Parameters}

Our numerical simulation is carried out with 40 quenched configurations
on a $\mbox{16}^{3} \times 48$ lattice separated by 1000 sweeps of 
the 5-hit Cabbibo-Marinari-Okawa algorithm\cite{Cabbibo_Marinari_Okawa}
at $\beta = \mbox{6.0}$.

The heavy quark masses used are $m_Qa=1000, 10.0, 7.0, 5.0, 4.0$
with the $n=1$ action and $5.0, 4.0, 3.0, 2.5$ with $n=2$.
The values $m_{Q}$=5.0 and 4.0 are used for both $n=1$ and 2 actions
in order to check consistency of results for the two actions.
The large value $m_{Q}=1000$ is taken to
compare with results of the static approximation.

For the light quark we used the Wilson action with
the hopping parameter $K=\mbox{0.1530}$, $\mbox{0.1540}$
and $\mbox{0.1550}$.
The critical hopping parameter is $K_{c}=0.15708(2)$.
The $\pi$ and $\rho$ meson masses and the pion decay constant
we obtained using the standard procedure
are given in Table \ref{light-light}.
The values of $m_{\rho}$ and $f_{\pi}$
extrapolated to $K \rightarrow K_{c}$
enables us to set the lattice spacing 
in terms of the physical values $m_{\rho}$=770 MeV
and $f_{\pi}$=132 MeV, yielding $a^{-1}$=2.3(3) GeV
consistently for both $m_{\rho}$ and $f_{\pi}$ where we used
$Z_{A}=0.86$\cite{Martinelli_Zhang}.

\subsection{Signal to Noise Ratio}

For each heavy and light quark masses
we have measured the local-local correlation function
\begin{equation}
C^{LL}(t) = \sum_{x} \langle 0 | T A_{4}(x) A_{4}^{\dagger}(0) | 0 \rangle
\end{equation}
where $A_{4}(x)$ is the heavy-light axial vector current given in
eq.(\ref{heavy-light_current}).
For large enough times this correlation function is dominated by
the ground state, i.e. the heavy-light meson of lowest mass,
\begin{equation}
C^{LL}(t) \longrightarrow \frac{(f_{P} m_{P})^{2}}{2 m_{P}} e^{-E t}
\label{asymptotic_behavior}
\end{equation}
where $E$ is the binding energy for the ground state.

\begin{figure}[h]
\hspace*{-0.5cm}
\epsfxsize=16cm
\epsffile{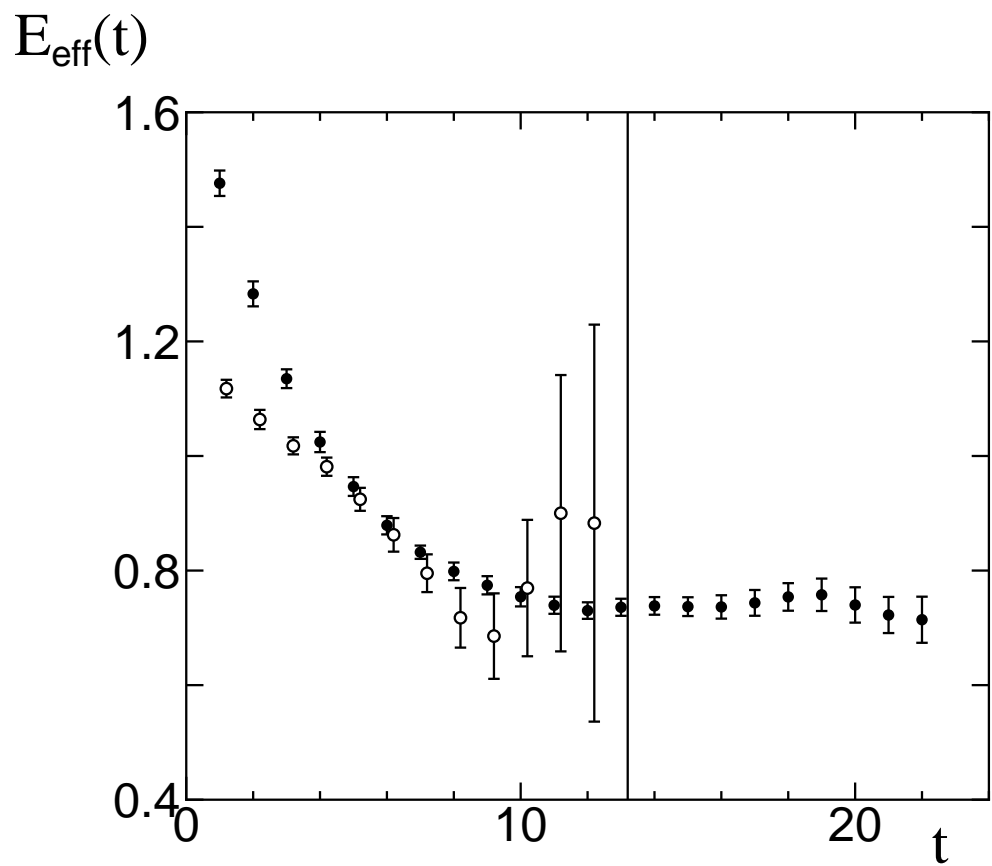}
\vspace{-1cm}
\caption{Effective binding energy  of the local-local correlation function
         for $m_{Q}=1000$ (open circles) and $m_{Q}=5$ (filled circles),
          both with $K=0.1530$ for light quark.}
\label{effective_mass}
\end{figure}

In Fig.\ref{effective_mass} we plot the effective binding energy defined by 
\begin{equation}
E_{eff}(t)=-\log(\frac{C(t+1)}{C(t)}), \quad C(t)=C^{LL}(t)
\end{equation}
for $m_{Q}$=1000 (open circles) and 5.0 (filled circles).
It is clear that the signal is far better for $m_{Q}$=5.0
for which we find a clear plateau at $t \geq$ 12.
It is well known
that the signal for the correlation function
in the static limit 
is very noisy, which we also observe here for $m_{Q}=1000$.

The improvement  of ground state signals for large but finite  values of
$m_Q$ can be qualitatively understood from the estimate of the relative
error\cite{Lepage}, \begin{equation}
\frac{\delta C(t)}{C(t)} \propto 
       \exp \left[ \big(E(Q\bar{q}) - \frac{E(Q\bar{Q}) + m_{\pi}}{2}\big) t
\right]
\label{relative_error}
\end{equation}
where $E(Q\bar{q})$ and $E(Q\bar{Q})$ are the binding energies
of heavy-light and heavy-heavy mesons respectively.
Values of binding energies for $K=0.1530$ are listed
in Table \ref{binding_energies:table}.  
For finite $m_Q$ the negative contribution from $E(Q\bar{Q})$
significantly reduces the value of the exponential slope
from that in the static limit where
$E(Q\bar{Q})=0$,
leading to a much milder growth of the relative error.

\begin{figure}[h]
\hspace*{-0.5cm}
\epsfxsize=16cm
\epsffile{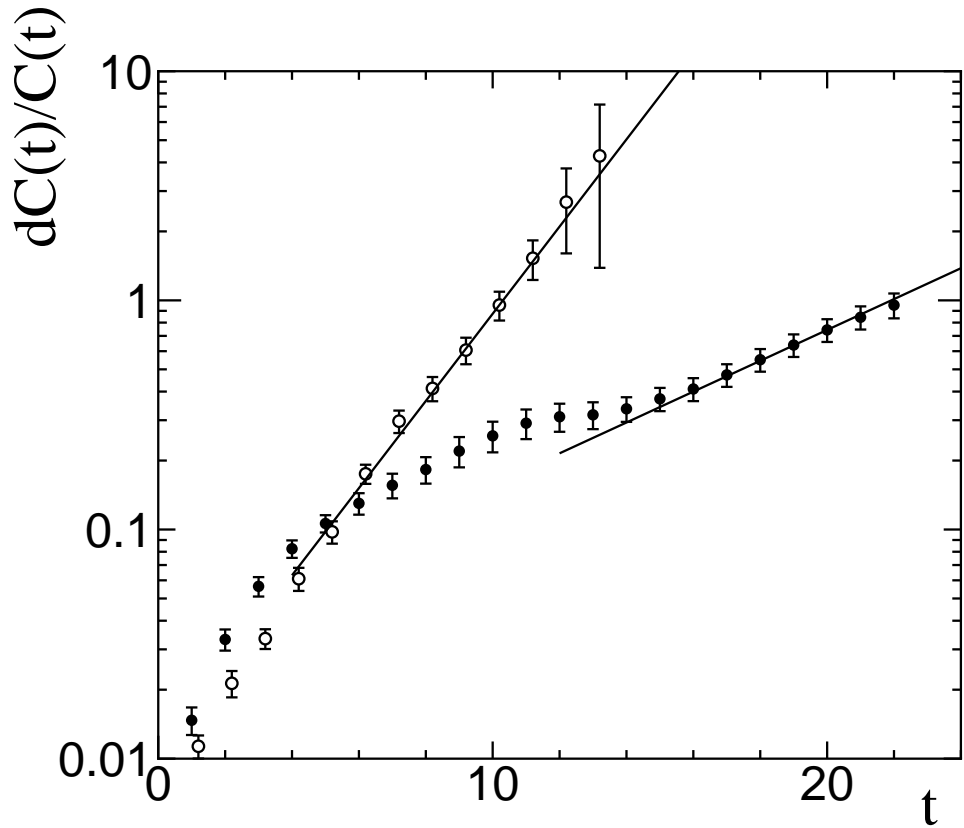}
\vspace{-1cm}
\caption{Relative error of local-local correlation function for $m_{Q}=1000$
         (open circles) and $m_{Q}=5$ (filled circles),
         both with $K=0.1530$ for light quark.
         Solid lines indicate the slope expected from
         measured values of the binding energies and $m_{\pi}$ according to
         (\protect\ref{relative_error}).
         }
\label{relative_error:figure}
\end{figure}

We found that our data for $\delta C(t)/C(t)$ are quantitatively
consistent with the above estimate.
In Fig.\ref{relative_error:figure} we show typical examples
of the relative error $\delta C(t)/C(t)$  
where solid lines indicate the slope expected from
the measured values of the binding energies and $m_{\pi}$ according to
(\protect\ref{relative_error}).
Fitting $\delta C(t)/C(t)$ with the exponential function
$\exp{(\alpha t)}$ we compare the results for $\alpha$ with
the estimate from the binding energies in Table \ref{binding_energies:table}.  
We observe quantitative agreement between $\alpha$ and 
$E(Q\bar{q}) - \frac{E(Q\bar{Q}) + m_{\pi}}{2}$ except for a few cases.

\subsection{Smearing}

In ref.\cite{Hashimoto_Saeki} we reported a disappointing fact that
the measured values of the heavy-light decay constant
in the static limit actually depend on the choice of smearing of the axial
vector current. Here we study if this problem is avoided in  
non-relativistic QCD.
We use the cube smearing as a typical choice of the
smearing function and examine the dependence of the decay constant
on the size of the cube.
The smeared current is defined as
\begin{equation}
A^{S}_{\mu}(x) \equiv \frac{1}{n^{3}}
               \sum_{i} ( Q^{\dagger}(x_{i}),  0 ) \gamma_{\mu} \gamma_{5}
                        q(x)
\end{equation}
where the sum is over points contained in a cube of a size $n^{3}$
centered at $x$.
We employ the Coulomb gauge fixing
instead of inserting gauge links between the heavy and light quark fields.
We compute the local-smeared (LS) and smeared-smeared (SS)
correlation functions defined by
\begin{eqnarray}
C^{LS}(t) & \equiv & \sum_{x} \langle 0 | T A^{L}_{4}(x,t) A^{S}_{4}(0,0)
                              | 0 \rangle \\
C^{SS}(t) & \equiv & \sum_{x} \langle 0 | T A^{S}_{4}(x,t) A^{S}_{4}(0,0)
                              | 0 \rangle
\end{eqnarray}
for the sizes of smearing $\mbox{3}^{3}$, $\mbox{5}^{3}$, $\mbox{7}^{3}$
and $\mbox{9}^{3}$.
Fitting these correlation functions as
\begin{eqnarray}
C^{LS}(t) & \longrightarrow & Z^{LS} \exp ( - E t ) \\
\frac{C^{SS}(t)}{C^{LS}(t)}
          & \longrightarrow & Z^{S/L}
\end{eqnarray}
for large enough $t$ regions,
we calculate the heavy-light decay constant using
\begin{equation}
\frac{1}{2} f_{P}^{2} m_{P} = Z_{A}^{2} \frac{Z^{LS}}{Z^{S/L}}
\end{equation}
where $Z_{A}$ is the renormalization factor of
the lattice axial vector current.

\begin{figure}[h]
\hspace*{-0.5cm}
\epsfxsize=16cm
\epsffile{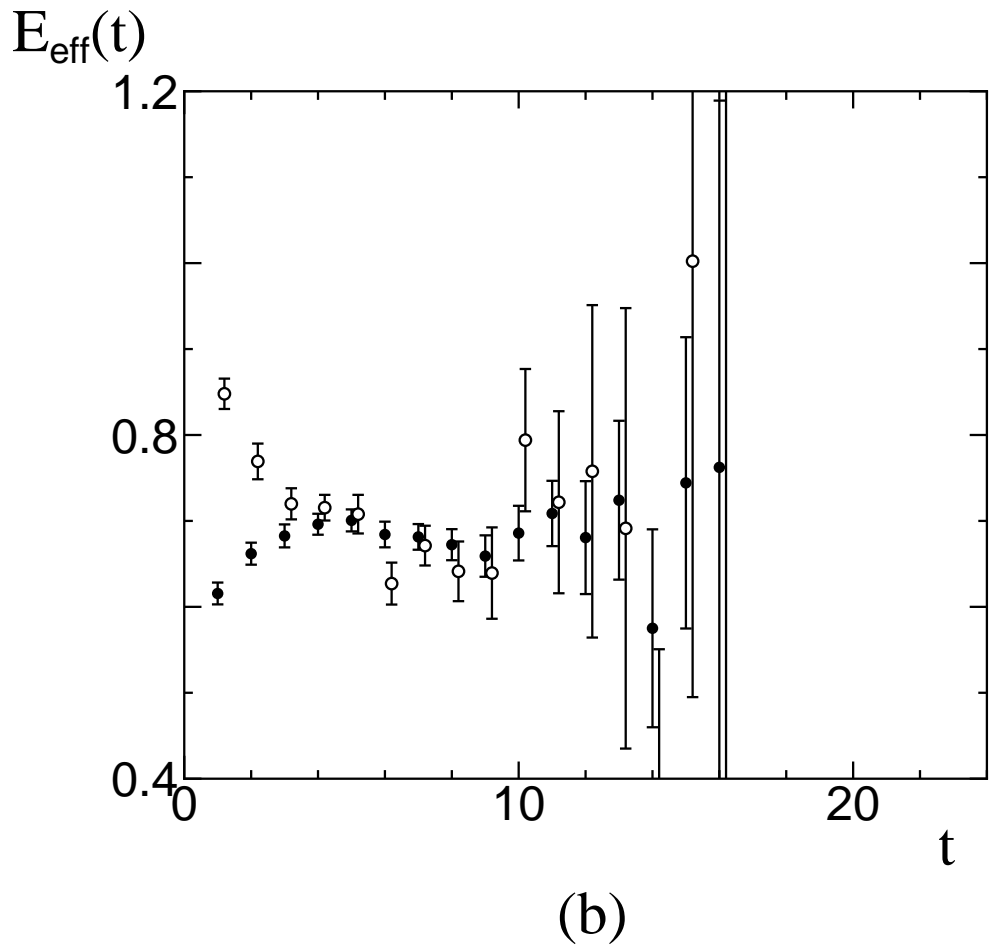}
\vspace{-1cm}
\caption{Effective binding energy of the local-smeared (LS) and
smeared-smeared(SS)
         correlation functions for (a) $m_{Q}$=5.0  and
         (b) $m_{Q}$=1000, both with $K$=0.1530 for light quark. Smearing size
is $5^3$.
         }
\label{effective_masses:smeared}
\end{figure}

In Fig.\ref{effective_masses:smeared} we plot
the effective binding energies of the LS(filled circles) and SS(open circles)
correlation functions with the $\mbox{5}^{3}$ smearing.
At $m_{Q}$=5.0 (Fig.\ref{effective_masses:smeared}(a))
clear signal of the ground state is observed beyond $t \approx$ 10
 both for LS and SS correlators.  Furthermore the values of the binding
energies are consistent. At $m_{Q}$=1000 
(Fig.\ref{effective_masses:smeared}(b)), on the other hand,
it is impossible to identify a plateau.

\begin{figure}[h]
\hspace*{-0.5cm}
\epsfxsize=16cm
\epsffile{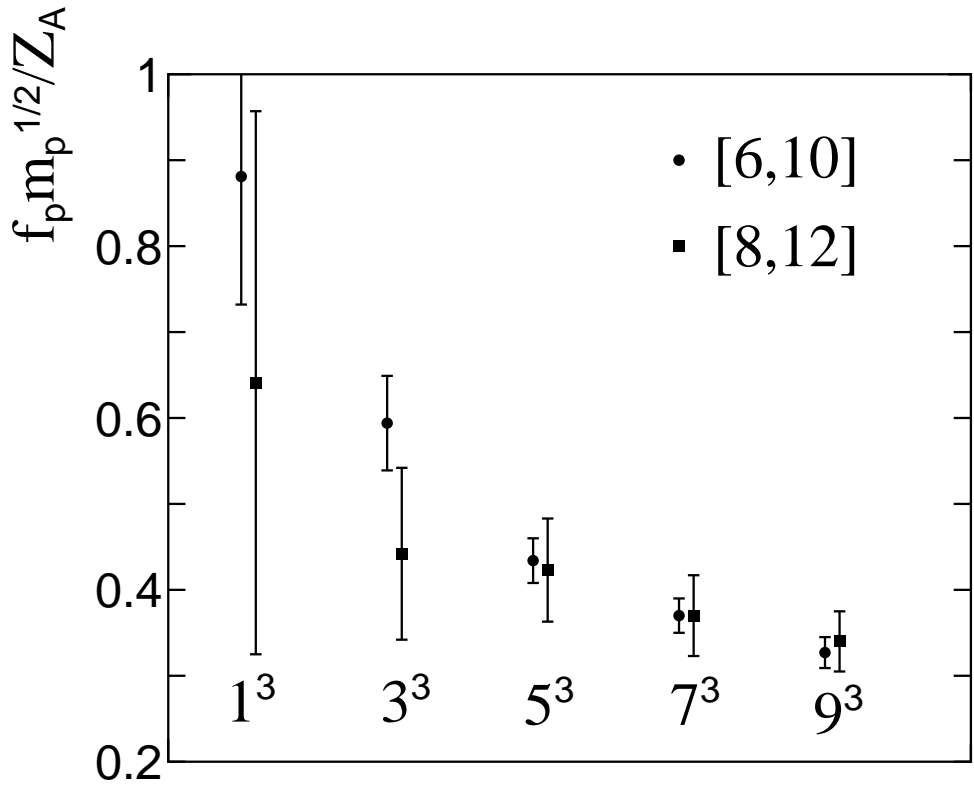}
\vspace{-1cm}
\caption{Dependence of $f_{P} \protect\sqrt{m_{P}}/Z_{A}$
         on the fitting range $t_{min} < t < t_{min}+\protect\mbox{4}$
         for various smearing sizes
         at $m_{Q}=$1000 (a) and $m_{Q}=$5.0 (b)
         with $K$=0.1530 for light quark.
         }
\label{smearing_dependence}
\end{figure}

In Fig.\ref{smearing_dependence} we plot the values of $f_{P} \sqrt{m_{P}} /
Z_{A}$ extracted from fits of the correlation functions
over 4 time slices $t_{min} \leq t \leq t_{max}=t_{min}+4$
for various smearing sizes.
For each group of data points $t_{min}$ is taken to be  
$t_{min}$=6, 8, 10 and 12 from left to right.
At $m_{Q}$=5.0 (Fig.\ref{smearing_dependence}(a))
we observe that the estimates converge to the same value
after $t_{min} \sim$ 10 for all the smearing sizes
including the case of no smearing (denoted by $1^3$ in
Fig.\ref{smearing_dependence}(a)). This gives us confidence that the asymptotic
region where the ground state dominates the correlation function 
is reached at $t \approx$ 10--12.
Furthermore it is interesting to notice that
the magnitude of errors are similar for various smearing sizes
including the case of no smearing.
This indicates  that the smearing technique
does not improve statistics for this case.

At $m_{Q}$=1000 (Fig.\ref{smearing_dependence}(b))
the situation is quite different.  Here only results for $t_{min}$=6 and 8 are 
available because of rapid growth of noise at $t >$ 10.
For these fitting intervals the data still depend on
the size of smearing, showing that the asymptotic region
is not yet reached.
It is essential for calculations in the static limit
to use some method which enhance ground state signals 
in the region $t <$ 10.
We do not use the data at $m_{Q}$=1000
in the following analysis.

\subsection{Decay Constant}

Because the results for the decay constant for $m_{Q} \leq$ 10.0
do not depend on the smearing size for $t_{min} \geq$ 10,
we choose the cube smearing of size $\mbox{5}^{3}$
and extract $f_{P} \sqrt{m_{P}}/Z_{A}$ from
a global fit over the interval 10 $\leq t \leq$ 20.
Other choices give similar results.

\begin{figure}[h]
\hspace*{-0.5cm}
\epsfxsize=16cm
\epsffile{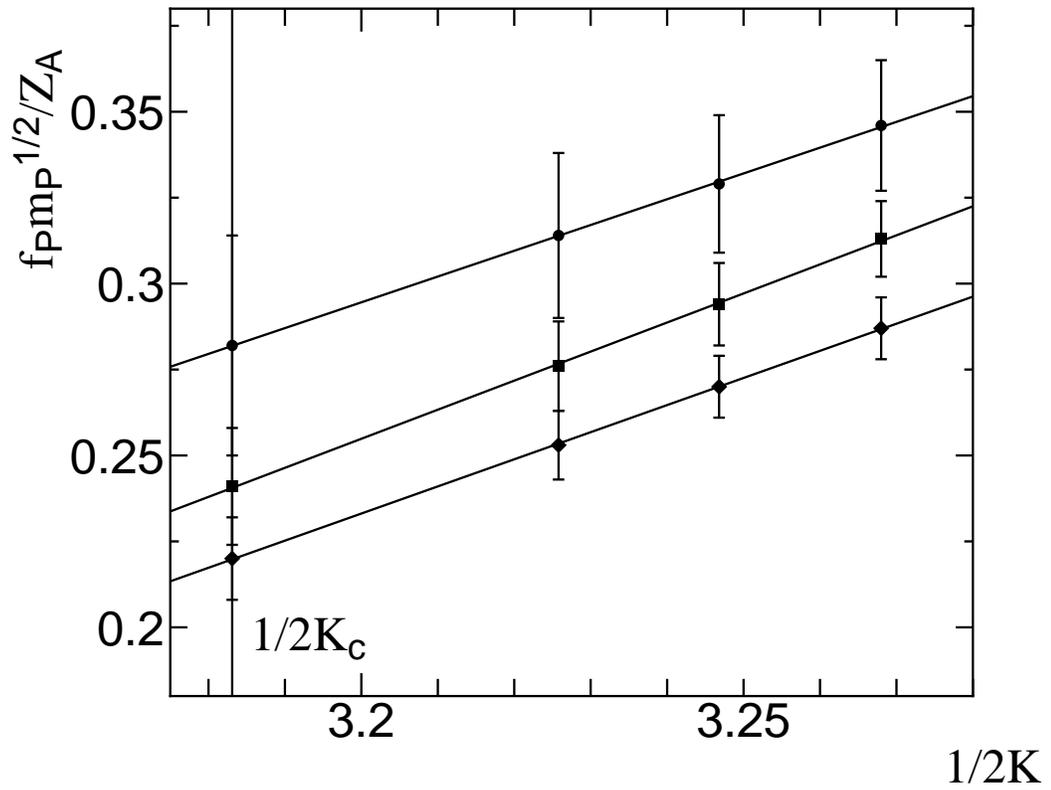}
\vspace{-1cm}
\caption{$f_{P} \protect\sqrt{m_{P}}/Z_{A}$ as a function
         of $1/2K$ for $m_{Q}$=10.0 (circles),
         5.0 (squares) and 2.5 (diamonds).
         The solid line indicates the extrapolation to
         the chiral limit $1/2K_{c}$.
        }
\label{1/K_dependence}
\end{figure}

In Table~\ref{data:table} we summarize our results for the 
binding energy $E$ and the 
decay constant $f_{P} \sqrt{m_{P}}/Z_{A}$
at each value of the hopping parameter of light quark $K$ and the heavy quark
mass $m_{Q}$.  Errors are estimated by the single elimination jackknife
procedure.   We observe that the two actions with $n=1$ and 2 yield consistent
values for  $f_{P} \sqrt{m_{P}}/Z_{A}$ for  $m_Q=4.0$ and 5.0 
where both actions are employed.
For each $m_{Q}$ we extrapolate the results at three values of $K$
linearly in $1/K$ to $1/K_{c}$ (see Fig.\ref{1/K_dependence}), the results of
which are also given in Table~\ref{data:table}.

\section{Extracting Physical Value of $f_B$}

In order to estimate the physical value of the $B$ meson
decay constant $f_{B}$
we have to determine the heavy quark mass $m_{Q}$ corresponding
to the $b$ quark and the renormalization factor of
the axial vector current $Z_{A}$.
A complete one-loop perturbative calculation necessary for this purpose is not
yet available for our non-relativistic QCD action
(\ref{action}) with $c$=1.  Davies and
Thacker\cite{Davies_Thacker_1,Davies_Thacker_2}, however, have reported the
one-loop results for the case of $c$=0, and we use their results incorporating
the  tadpole improvement of lattice perturbation theory
\cite{Lepage_Mackenzie}.  

\subsection{Improved Perturbation Theory for Non-relativistic QCD}

To one-loop order the inverse heavy quark propagator
$\Delta_{Q}$ can be written in the form,
\begin{equation}
\Delta_{Q} =
 (1-Cg^{2}) \left[ (e^{i p_{0}} -1-Ag^{2})
                   +(1-Bg^{2}) \frac{\mbox{\bf p}^{2}}{2 m_{Q}}
            \right]
\end{equation}
where $A$, $B$ and $C$ are numerical constants.
We define the mass renormalizaion factor $Z_{m}$,
the energy shift $E_{0}$ and
the wave function renormalization factor $Z_{Q}$ by
\begin{eqnarray}
G^{pert}(\mbox{\bf p},t) & = &
 \int^{+\pi}_{-\pi} \frac{dp_{0}}{2\pi} \frac{1}{\Delta_{Q}}
 e^{i p_{0} t} \\ \nonumber
 & = & \frac{1}{1-Cg^{2}} \left[
       (1+Ag^{2})-(1-Bg^{2}) \frac{\mbox{\bf p}^{2}}{2 m_{Q}}
       \right]^{t-1} \\ \nonumber
 & \equiv & Z_{Q}^{pert} \exp \left[ - t
            \left( E_{0}^{pert} +
            \frac{\mbox{\bf p}^{2}}{2 Z_{m}^{pert} m_{Q}}
            \right) \right].
\end{eqnarray}
We then obtain
\begin{eqnarray}
E_{0}^{pert} & = & - \ln [1 + A g^{2}] \nonumber \\
Z_{m}^{pert} & = & 1 + (A + B) g^{2} \nonumber \\
Z_{Q}^{pert} & = & \frac{1}{1 - (C - A) g^{2}}.
\label{pert}
\end{eqnarray}
For these quantities the mean-field improved\cite{Lepage_Mackenzie} 
expressions are given by
\begin{eqnarray}
E_{0}^{mf} & = & - \ln \left[ u_{0}
                       \left( 1 - \frac{3}{n m_{Q}} (1-u_{0})
                       \right)^n \right] \nonumber \\
Z_{m}^{mf} & = & \frac{1}{u_{0}} 
                       \left( 1 - \frac{3}{n m_{Q}} (1-u_{0})
                       \right) \nonumber \\
Z_{Q}^{mf} & = & \frac{1}{\left(1 - \frac{3}{n m_{Q}} (1-u_{0})\right)^{n}}
\label{mf}
\end{eqnarray}
where $u_{0}$ is a mean-field value of the lattice link
variable $U_{\mu}(x)$.
As a definition of $u_{0}$ we take the simplest choice
$u_{0}=\langle \frac{1}{3} \mbox{Tr} U_{plaq} \rangle^{1/4}$
for which the perturbative expansion is given by
$u_{0}^{pert} = 1-\frac{1}{12} g^{2}$.

Tadpole-improved one-loop results for the renormalization factors are
obtained by combining (\ref{mf}) and (\ref{pert})
after removing one-loop terms
of the mean-field expressions (\ref{mf}) from $A, B$ and $C$.
This gives 
\begin{eqnarray}
E_{0}^{tad-imp} & = & - \ln \left[ u_{0}
                        \left( 1 - \frac{3}{n m_{Q}} (1-u_{0})
                        \right)^{n}
                        \left\{1 + \tilde{A} g^{2} \right\}
                        \right] \nonumber \\
Z_{m}^{tad-imp} & = & \frac{1}{u_{0}} 
                        \left( 1 - \frac{3}{n m_{Q}} (1-u_{0})
                        \right)
                        \left\{1 + \widetilde{(A+B)} g^{2}
                        \right\} \nonumber \\
Z_{Q}^{tad-imp} & = & \frac{1}{
                       \left(1 - 
                       \frac{3}{n m_{Q}} (1-u_{0}) \right)^{n} }
                       \left\{1 + \widetilde{(C-A)} g^{2} \right\}.
\end{eqnarray}
where $\tilde{A}$, $\widetilde{(A+B)}$, $\widetilde{(C-A)}$ are again
numerical constants defined by
\begin{eqnarray}
\tilde{A}         & = & A + \frac{1}{12} 
                        \left(1+\frac{3}{m_{Q}} \right) \nonumber \\
\widetilde{(A+B)} & = & (A+B) - \frac{1}{12}
                        \left(1-\frac{3}{n m_{Q}} \right) \nonumber \\
\widetilde{(C-A)} & = & (C-A) - \frac{1}{4 m_{Q}}.
\end{eqnarray}
In Table \ref{ren_params} we summarize numerical values
of these quantities for some typical values of $m_{Q}$
where we use the values obtained by Davies and Thacker
for the numerical coefficients $A$, $B$ and $C$.
Smallness of the `tilde' quantities compared with the original
values demonstrates that the tadpole improvement works well
for these quantities.

The values of $E_0, Z_m$ and $Z_Q$ are also tabulated in Table
\ref{ren_params}. For the coupling constant $g^{2}$ we take
$g_{V}^{2}(\pi/a)=1.96$ at $\beta=6.0$\cite{Lepage_Mackenzie}.    In principle
it is more desirable to use $g_{V}^{2}(q^{\ast})$ with a properly determined
scale $q^{\ast}$ as proposed in \cite{Lepage_Mackenzie}.
Morningstar has calculated this scale $q^{\ast}$ for
a more complicated non-relativistic QCD action\cite{Morningstar}
and obtained $q^{\ast}$=0.67$a^{-1}$ for $Z_{m}$
and 0.81$a^{-1}$ for $E_{0}$.  The corresponding values of
$g_{V}^{2}(q^{\ast})$ ({it e.g.,} $g_{V}^{2}(0.67/a)=3.41$) are substantially
larger than $g_{V}^{2}(\pi/a)$.  Because of the smallness of the
tadpole-improved one-loop coefficients, however, the
renormalization constants are modified only slightly; the combination
$Z_mm_Q-E_0$ which is relevant for the $B$ meson mass (see below) changes only by
2--3\%, and the change of the heavy quark wave function renormalization
factor $Z_Q$ is also at the level of a few percent.

\subsection{$b$ Quark Mass}

\begin{figure}[h]
\hspace*{-0.5cm}
\epsfxsize=16cm
\epsffile{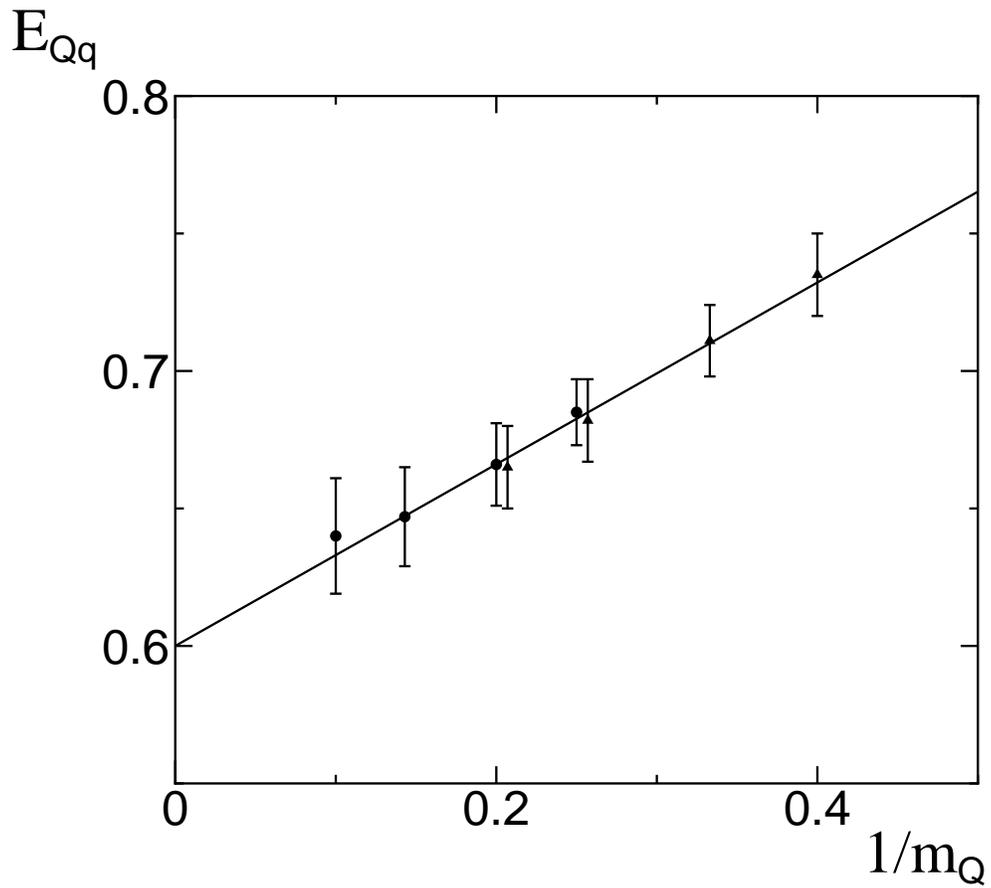}
\vspace{-1cm}
\caption{The binding energy $E_{Q\bar{q}}$ of
         the heavy-light meson as a function of $1/m_{Q}$
         Circles and triangles are for the results
         obtained with the $n$=1 and $n$=2 actions respectively.
         The solid line represents the fit
         $E_{Q\bar{q}}=\mbox{0.60}+\mbox{0.33}/m_{Q}$.
         }
\label{bene:figure}
\end{figure}

The heavy-light meson mass is given by
\begin{equation}
m_{P} = Z_{m} m_{Q} - E_{0} + E_{Q\bar{q}}.
\end{equation}
In Fig.~\ref{bene:figure} we plot our numerical result for the
binding energy $E_{Q\bar{q}}$ as a function of $1/m_Q$.  The solid line
$E_{Q\bar{q}}=\mbox{0.60}+\mbox{0.33}/m_{Q}$ fits the data very well.
Using this fit and the perturbative result for $Z_m$ and $E_0$ discussed
in Sec.~4.1, together with the inverse lattice spacing $a^{-1}$ = 2.3(3) GeV
obtained from the $\rho$ meson mass, we find $m_{P}$ = 12.6(1.6), 7.2(9) and
5.4(7) GeV for $m_{Q}$ = 5.0, 2.5 and 1.8 respectively.  This shows that the
heavy quark mass of  $m_{Q}$ = 1.8 in lattice units
approximately corresponds to the $b$ quark.
 This value is consistent with the estimate given by Davies and Thacker
using $\Upsilon$  spectroscopy\cite{Davies_Thacker_3}.
In the following analysis of the $B$ meson decay constant
we use $m_{Q}$ = 1.8 for the $b$ quark mass.

\subsection{Axial vector current renormalization factor $Z_{A}$}

The renormalization factor $Z_{A}$ for the axial vector current can be
written as \begin{equation}
Z_{A} = \frac{Z_{A}^{cont}}{Z_{A}^{latt}}.
\end{equation}
Here $Z_{A}^{cont}$ is the renormalization constant
for the continuum axial vector current renormalized with
the $\overline{\mbox{MS}}$ scheme\cite{Eichten_Hill} given by
\begin{equation}
Z_{A}^{cont} = 1 + \frac{1}{12 \pi^{2}} g^{2}\left(
\frac{3}{2} \log (m_{Q}^{cont})^2 - \frac{3}{4} \right)
\label{Z_A_cont}
\end{equation}
where we use $m_{Q}^{cont}=Z_{m} m_{Q}$ as the heavy quark mass.
We ignore infra-red divergent logarithms since it cancels
with the lattice counterpart.
The lattice renormalization factor $Z_{A}^{latt}$ can be written as
\begin{equation}
Z_{A}^{latt} = Z_{Q}^{1/2} Z_{q}^{1/2} [1 + V^{latt} g^{2}]
\end{equation}
where $Z_{Q}$ is the wave function renormalization factor 
for the heavy quark discussed in Sec.~4.1.  
For the Wilson light quark our field normalization includes the conventional
factor $\sqrt{2K_c}$.  The tadpole-improved one-loop result for the wave function
renormalization factor $Z_{q}$  is then given
by\cite{Lepage_Mackenzie} \begin{equation}
Z_{q} = \frac{2 K_{c}}{1/4} [ 1 + 0.0043 g^{2} ].
\end{equation}
Finally the vertex correction $V^{latt}$ is not
affected by the tadpole improvement. 
We use the value given in ref.~\cite{Davies_Thacker_2}
for the spinless non-relativistic QCD action ($c$=0) since theirs is the only
result available for this quantity at present. Combining
these quantities we obtain the final expression, \begin{eqnarray}
Z_{A} & = & 
        \left( 1-\frac{3}{n m_{Q}} (1-u_{0}) \right)^{n/2}
        \left( \frac{1/4}{2 \kappa_{c}} \right)^{1/2} \\ \nonumber
      &   &
        \times \left[ 1 - \left( \tilde{D}-\frac{1}{4 \pi^{2}} 
                                 \ln (m_{Q}^{cont} a)
                          \right) g^{2}
        \right]
\end{eqnarray}
where $\tilde{D}$ is a numerical constant.  

In  Table \ref{Z} we list representative values of
$V^{latt}, \tilde D, 1/Z_{A}^{latt}$ and $Z_A$.
It is interesting to note that the value of $Z_{A}^{latt}$ is
almost independent of $1/m_{Q}$
while $Z_{A}$ shows logarithmistic dependence 
and even diverges at $m_{Q}=\infty$
reflecting the form of $Z_{A}^{cont}$ (eq.(\ref{Z_A_cont})).
We also observe that the expansion coefficient $\tilde{D}$ is still large
after the tadpole improvement,
suggesting the possibility that higher order coefficients may not be
small\cite{Heavy-Light_review}. The main contribution to the large coefficient
is the vertex correction $V^{latt}$
which is unmodified by the tadpole improvement.
This is a general feature for renormalization constants
of bilinear operators of light-light, heavy-light and heavy-heavy quarks.

The large one-loop coefficient increases the magnitude of
uncertainties coming from the choice of the scale $q^{\ast}$ in the
coupling constant $g_{V}^{2}(q^{\ast})$.
In the static limit, the optimal value of $q^{\ast}$
is estimated\cite{Hernandez_Hill} as $q^{\ast}=2.18/a$
for which $g_{V}^{2}(q^{\ast})$=2.22 at  $\beta=6.0$.
This choice of $q^{\ast}$ leads to $Z_{A}$=0.661 at $m_Q=\infty$,
while another choice $q^{\ast}=1/a$ for which
$g_{V}^{2}(q^{\ast})$=3.10 gives $Z_{A}$=0.570
which is 14 \% smaller.
In the following we use $q^{\ast}=2.18/a$ for a
reference value,
keeping in mind this large systematic uncertainty.

\subsection{Decay Constant}

\begin{figure}[h]
\hspace*{-0.5cm}
\epsfxsize=16cm
\epsffile{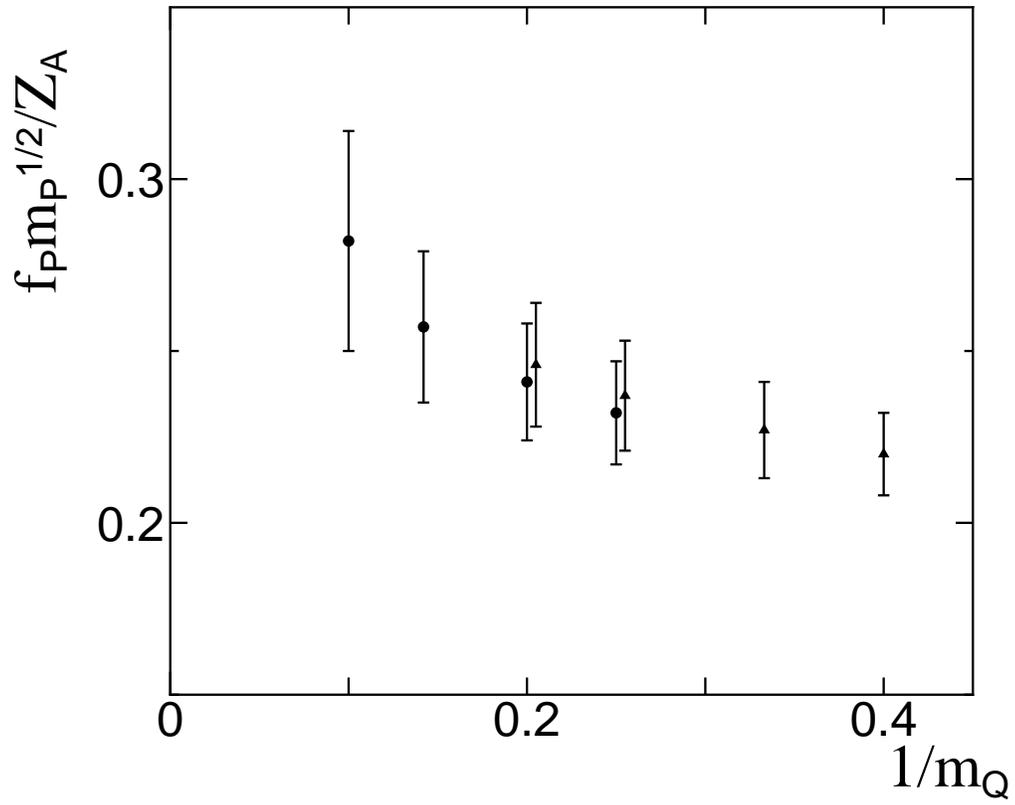}
\vspace{-1cm}
\caption{$1/m_{Q}$ dependence of
         $f_{P} \protect\sqrt{m_{P}}/Z_{A}$ after
         the extrapolation to the limit $K=K_{c}$
         for light quark.
         Circles and triangles are for the results
         obtained with the $n$=1 and $n$=2 actions respectively.
         }
\label{data:figure}
\end{figure}

Our results for $f_{P} \sqrt{m_{P}}/Z_{A}$ after extrapolation to $K=K_c$ for
light quark are plotted as a function of $1/m_{Q}$ in Fig.\ref{data:figure}
(see Table~\ref{data:table} for numerical values). Circles and triangles are for
the  results obtained with the $n=1$ and $2$ actions respectively.
The two actions yield consistent values 
for  $m_Q=4.0$ and 5.0  as we already noted in Sec.~3.4.

We observe in Fig.~\ref{data:figure} a clear deviation from the
scaling law of the heavy-light decay constant in the heavy mass limit given by 
\begin{equation}
f_{P} \sqrt{m_{P}}/Z_{A} = \mbox{const}.
\end{equation}
In order to evaluate the magnitude of the  deviation
we fit the data with the form
\begin{equation}
f_{P} \sqrt{m_{P}}/Z_{A} = (f_{P} \sqrt{m_{P}}/Z_{A})^{static}
                           \left( 1 - \frac{c}{m_{Q}} \right).
\label{fit:equation}
\end{equation}
In order to estimate systematic uncertainties
in the fit due to a slight curvature in the $1/m_Q$ dependence of $f_{P}
\sqrt{m_{P}}/Z_{A}$, we list in Table~\ref{m-inv-fit} the results of fit for
four representative selection of data points.  The first two choices 
employ all data points except at $m_Q=4.0$ and 5.0 where the results with the
$n=2$ action is chosen for the choice (a) and those with the $n=1$ action for
the choice (b).  The choice (c) uses data for $5.0\geq m_Q\geq 2.5$ with
the $n=2$ action and (d) for $10.0\geq m_Q\geq 4.0$ with $n=1$.  Errors given
in Table~\ref{m-inv-fit} are estimated by a single elimination jackknife
procedure. 

For the decay constant in the static limit
$(f_{P} \sqrt{m_{P}}/Z_{A})^{static}$
all four choices yield values consistent
within 10--15\%.  As a representative value we
quote the result for the choice (a);
\begin{equation} (f_{P}
\sqrt{m_{P}}/Z_{A})^{static} = \mbox{0.292(35),\hspace{10mm}}
c = \mbox{1.04(44)}
\label{fit:result}
\end{equation}

The extrapolated value (\ref{fit:result}) is compared 
with the results of other groups obtained at $\beta=6.0$ using the static heavy
quark propagator in Table \ref{comparison:data}.  The results are
consistent in view of the systematic uncertainties in the static results
depending on the detail of smearing.

Using the value (\ref{fit:result}) for $(f_{P} \sqrt{m_{P}}/Z_{A})^{static}$ 
we obtain the heavy-light decay constant in the static limit, 
\begin{equation}
(f_{P} \sqrt{m_{P}})^{static} = 
              0.682\pm 82^{+34}_{-70}
              \left( \frac{Z_{A}}{0.67} \right)
              \left( \frac{a^{-1}}{2.3 \mbox{GeV}} \right)^{3/2}
              \mbox{GeV}^{3/2}
\end{equation}
where the first error is statistical
and the second one is systematic reflecting variation of central values among
the four selection of data points in Table~\ref{m-inv-fit}.
For the renormalization factor of the axial current $Z_{A}$
which diverges in this limit
we use the value evaluated at the $b$-quark mass as a normalization
as usual.
For the $B$ meson decay constant evaluated in the static
limit this yields
\begin{equation}
f_{B}^{static} = 297\pm 36^{+15}_{-30}
              \left( \frac{Z_{A}}{0.67} \right)
              \left( \frac{a^{-1}}{2.3 \mbox{GeV}} \right)
              \mbox{MeV}.
\end{equation}

In order to obtain the $B$ meson decay constant including $1/m_Q$
corrections we extrapolate the fit (\ref{fit:equation}) to the $b$ quark mass
of $m_{Q}$=1.8 as was discussed in Sec.4.2, employing the single elimination
jackknife procedure for estimating the statistical error of the extrapolated
value.   The result shows a sizable variation depending on the selection
of data points.  We quote the value for the fitting  choice (d) in Table
\ref{m-inv-fit} employing data for $5.0\geq m_Q\geq 2.5$, which are close
to the extrapolated point $m_Q=1.8$ and hence should be more
reliable;
\begin{equation} f_{B} = 171\pm 22^{+19}_{-45}
                \left( \frac{Z_{A}}{0.67} \right)
                \left( \frac{a^{-1}}{2.3 \mbox{GeV}} \right)
                \mbox{MeV}.
\label{eq:fb}
\end{equation}
where the first error is statistical
and the second one is systematic showing the 
uncertainty originating from the choice of data points for the fitting.
It should be remembered that a systematic uncertainty also exists in the
determination of $Z_{A}$ and $a^{-1}$. 

The UKQCD collaboration\cite{UKQCD_nr-light} recently reported a preliminary
non-relativistic QCD result for $f_B$ obtained in a simulation carried out
at $m_Q=1.7$ on a $16^3\times 48$ lattice.  Their result
$f_B\sqrt{m_B}/Z_A=0.16(3)$ is consistent with our value obtained by an
extrapolation in $1/m_Q$.

\begin{figure}[h]
\hspace*{-0.5cm}
\epsfxsize=16cm
\epsffile{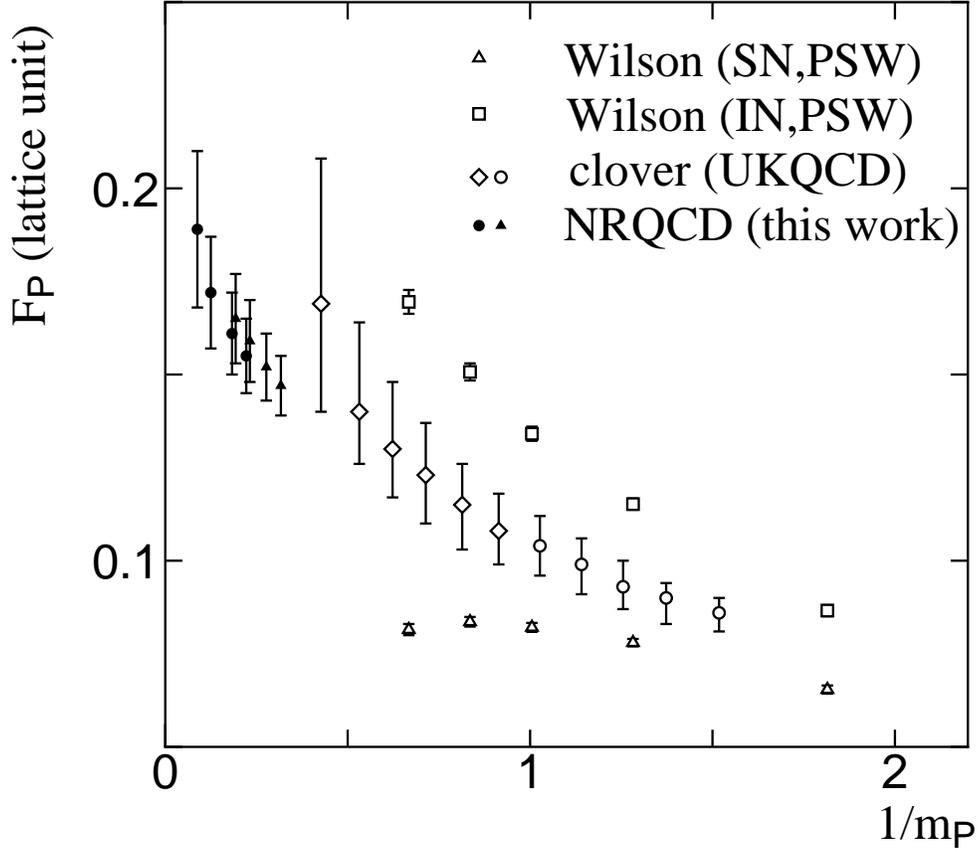}
\vspace{-1cm}
\caption{$F_{P} = (\protect\alpha_{s}(m_{P})/\protect\alpha_{s}(m_{B}))^{2/11}
         f_{P} \protect\sqrt{m_{P}}$
         as a function of $1/m_{P}$ after
         the extrapolation to the limit $K=K_{c}$
         for light quark.
         Open squares and tringles are for results of
         the PSI-Wuppertal collaboration\protect\cite{PSW_Wilson}
         using the Wilson action with the standard normalization
         $\protect\sqrt{2K}$ (triangles) and with the improved normalization
         $\protect\sqrt{1-3K/4K_{c}}$ (squares).
         Open circles and diamonds are for results of 
         the UKQCD collaboration\protect\cite{UKQCD_heavy-light}
         using the $O(a)$-improved (clover) fermion action.
         Closed symbols are for results of this work.
         }
\label{compare}
\end{figure}

It is also interesting to compare our results
with the results obtained using propagating quark for the heavy quark.
Two groups have reported the results of $f_{P} \sqrt{m_{P}}$
obtained using the propagating quarks at $\beta=$6.0.
In Fig.\ref{compare} we plot the quantity
\begin{equation}
F_{P} = \left( \frac{\alpha_{s}(m_{P})}{\alpha_{s}(m_{B})}
        \right)^{2/11} f_{P} \sqrt{m_{P}}
\end{equation}
as a function of $1/m_{P}$.
The normalization factor
$(\alpha_{s}(m_{P})/\alpha_{s}(m_{B}))^{2/11}$
is introduced in order to
absorb the logarithmistic divergence of $f_{P} \sqrt{m_{P}}$
at $m_{P}=\infty$ originating from $Z_{A}^{cont}$ (eq.(\ref{Z_A_cont})).
As a coupling constant we use $\alpha_{V}$ with $\Lambda_{V}$=0.169
at $\beta$=6.0.
Open squares and tringles are for results of
the PSI-Wuppertal collaboration\cite{PSW_Wilson}
using the Wilson action with the standard normalization
$\sqrt{2K}$ (triangles) and with the improved normalization
$\sqrt{1-3K/4K_{c}}$ (squares).
Open circles and diamonds are for results of 
the UKQCD collaboration\cite{UKQCD_heavy-light}
using the $O(a)$-improved (clover) fermion action.
Closed symbols are for results of this work.
We can see that our results are consistent with the results of
the clover action in view of the $1/m_{P}$ dependence of $F_{P}$.
For the Wilson action there is a source of large systematic uncertainty
in the choice of the normalization for this heavy mass region
and our results are not seen to be consistent with both
of these choices.

\section{Conclusions}

In this article we have reported on a calculation of the heavy-light decay
constant using non-relativistic lattice QCD.
We found that ground state signals in  the correlator is significantly
improved compared to those in the static limit, and that 
the degree of improvement of signal to noise ratio is in a quantitative
agreement with the estimate of  Lepage in terms of binding energies.
As a result we could extract properties of the ground state reliably.
In particular an apparent dependence of the decay constant
on the size of smearing for source,
which affected previous attempts in the static limit, is absent.

Our result for the $B$ meson decay constant shows that
the $1/m_Q$ correction to the static limit is quite significant 
even for the $b$ quark.
This points toward the necessity of a more complete calculation to
order $1/m_Q$ than was attempted here, and eventually a calculation
including $1/m_Q^2$ terms, for a precise determination of the $B$ meson decay
constant.  The improvements of the present work needed to order $1/m_Q$ are
the inclusion of $1/m_Q$ terms in the axial vector current and one-loop
calculation of renormalization factors including the spin-magnetic
interaction.  It is also desirable to estimate two-loop vertex corrections in
view of the large one-loop coefficient. We leave these problems for future
investigations.

\section*{Acknowledgements}

I am grateful to A.~Ukawa for many discussions and
a critical reading of the manuscript.
I would also like to thank A.~Kronfeld, P.~Mackenzie and M.~Okawa
for useful discussions.
The numerical computation were performed on HITAC S820/80 at KEK.
I am grateful to the Theory Division of KEK for warm hospitality.
The author was supported in part by the Grant-in-Aid of the
Ministry of Education under the contract No. 040011.

\newpage

\begin{table}
\begin{center}
\begin{tabular}{|c|ccc|}
\hline
$K$     & $m_{\pi}$ & $m_{\rho}$ & $f_{\pi}/Z_{A}$ \\
\hline
0.1530  & 0.419(4)  & 0.502(12)  & 0.105(4) \\
0.1540  & 0.360(5)  & 0.461(14)  & 0.096(5) \\
0.1550  & 0.291(7)  & 0.416(21)  & 0.081(6) \\
\hline
$K_{c}$ & 0         & 0.338(40)  & 0.067(11) \\
\hline
\end{tabular}
\end{center}
\caption{$\pi$ and $\rho$ meson masses and the pion decay constant $f_\pi$ 
         devided by the renormalization constant of the axial current
         in lattice units.
         The values of $m_{\rho}$ and $f_{\pi}/Z_{A}$ at
         the critical hopping parameter are obtained by a linear
extrapolation in $1/K$.} \label{light-light}
\end{table}
\begin{table}
\begin{center}
\begin{tabular}{|l|l|l|c|c|}
\hline
$m_{Q}$  &  $E(Q\bar{q})$ & $E(Q\bar{Q})$ &
$E(Q\bar{q}) - \frac{E(Q\bar{Q}) + m_{\pi}}{2}$ &
$\alpha$ \\ \hline
\hline
1000  &  0.647(18)  &  0          &   0.438(18) & 0.466(10)\\
10.0  &  0.713(24)  &  0.510(08)  &   0.249(24) & 0.155(51)\\
 7.0  &  0.720(16)  &  0.638(12)  &   0.192(17) & 0.153(30)\\
 5.0($n$=1) & 0.740(14) & 0.751(12) & 0.155(15) & 0.149(22)\\
 5.0($n$=2) & 0.738(13) & 0.729(13) & 0.164(15) & 0.147(24)\\
 4.0($n$=1) & 0.759(14) & 0.827(11) & 0.136(15) & 0.125(22)\\
 4.0($n$=2) & 0.756(12) & 0.804(09) & 0.145(13) & 0.138(22)\\
 3.0  &  0.786(12)  &  0.906(14)  &   0.124(14) & 0.116(23)\\
 2.5  &  0.810(15)  &  0.977(14)  &   0.112(17) & 0.101(23)\\
\hline
\end{tabular}
\end{center}
\caption{The heavy-light and heavy-heavy binding energies
         $E(Q\bar{q})$ and $E(Q\bar{Q})$ at $K$=0.1530 for
         the light quark.
         The combination
         $E(Q\bar{q}) - \frac{E(Q\bar{Q}) + m_{\pi}}{2}$
         determines the behavior of the noise
         (see  Eq.(\protect\ref{relative_error}) in text).  
         $\alpha$ denotes the measured value of the 
         exponential slope.
         }
\label{binding_energies:table}
\end{table}
\begin{table}
\begin{center}
\begin{tabular}{|c|c|c|c|c|}
\hline
\multicolumn{5}{|l|}{Binding energy $E$} \\ \hline
\makebox[12mm]{$m_{Q}$}                 & 
          \makebox[25mm]{$K$=    0.1530} &
          \makebox[25mm]{        0.1540} &
          \makebox[25mm]{        0.1550} &
          \makebox[25mm]{$K_{c}$=0.157} \\ \hline
  10.0  & 0.696(15) & 0.681(15) & 0.668(17) & 0.640(21) \\
   7.0  & 0.711(14) & 0.695(13) & 0.679(16) & 0.647(18) \\
   5.0  &           &           &           &           \\
($n$=1) & 0.734(12) & 0.717(14) & 0.700(14) & 0.666(15) \\
($n$=2) & 0.732(12) & 0.714(14) & 0.698(14) & 0.665(15) \\
   4.0  &           &           &           &           \\
($n$=1) & 0.755(13) & 0.737(12) & 0.720(14) & 0.685(12) \\
($n$=2) & 0.750(12) & 0.733(15) & 0.716(13) & 0.682(15) \\
   3.0  & 0.781(15) & 0.763(14) & 0.746(13) & 0.711(13) \\
   2.5  & 0.806(13) & 0.788(13) & 0.770(12) & 0.735(15) \\ \hline
\end{tabular}

\begin{tabular}{|c|c|c|c|c|}
\hline
\multicolumn{5}{|l|}{Decay constant $f_{P} \sqrt{m_{P}}/Z_{A}$} \\ \hline
\makebox[12mm]{$m_{Q}$}                 & 
          \makebox[25mm]{        0.1530} &
          \makebox[25mm]{        0.1540} &
          \makebox[25mm]{        0.1550} &
          \makebox[25mm]{$K_{c}$=0.157} \\ \hline
  10.0  & 0.346(19) & 0.329(20) & 0.314(24) & 0.282(32) \\
   7.0  & 0.328(14) & 0.310(15) & 0.292(17) & 0.257(22) \\
   5.0  &           &           &           &           \\
($n$=1) & 0.313(11) & 0.294(12) & 0.276(13) & 0.241(17) \\
($n$=2) & 0.319(12) & 0.300(13) & 0.282(14) & 0.246(18) \\
   4.0  &           &           &           &           \\
($n$=1) & 0.302(10) & 0.284(11) & 0.266(12) & 0.232(15) \\
($n$=2) & 0.309(11) & 0.291(11) & 0.272(13) & 0.237(16) \\
   3.0  & 0.295(09) & 0.278(10) & 0.261(11) & 0.227(14) \\
   2.5  & 0.287(09) & 0.270(09) & 0.253(10) & 0.220(12) \\ \hline
\end{tabular}
\end{center}
\caption{
         Binding energy $E$ and
         decay constant $f_{P} \protect\sqrt{m_{P}}/Z_{A}$
         at $K$=0.1530, 0.1540, 0.1550.  Values 
         at $K_{c}$=0.15708(2) are obtained by
         a linear extrapolation in $1/K$.
         }
\label{data:table}
\end{table}
\begin{table}
\begin{center}
\begin{tabular}{|c|c|c|c|c|c|}
\hline
\multicolumn{6}{|l|}{$E_{0}$} \\ \hline
$m_{Q}$ & \makebox[15mm]{$A$}            &
          \makebox[15mm]{$\widetilde{A}$}    &
          \makebox[15mm]{$E_{0}^{pert}$} &
          \makebox[15mm]{$E_{0}^{mf}$}   &
          \makebox[15mm]{$E_{0}^{tad-imp}$} \\ \hline
$\infty$ & -0.1684 & -0.0851 & 0.401 & 0.129 & 0.310 \\
5.0      & -0.2075 & -0.0742 & 0.522 & 0.202 & 0.359 \\
2.5      & -0.2487 & -0.0654 & 0.668 & 0.277 & 0.414 \\
1.8      & -0.2809 & -0.0586 & 0.800 & 0.338 & 0.460 \\ \hline
\end{tabular}
\begin{tabular}{|c|c|c|c|c|c|}
\hline
\multicolumn{6}{|l|}{$Z_{m}$} \\ \hline
$m_{Q}$ & \makebox[15mm]{$A+B$}            &
          \makebox[15mm]{$\widetilde{A+B}$}    &
          \makebox[15mm]{$Z_{m}^{pert}$} &
          \makebox[15mm]{$Z_{m}^{mf}$}   &
          \makebox[15mm]{$Z_{m}^{tad-imp}$} \\ \hline
$\infty$ &  0.0686 & -0.0147 & 1.13 & 1.14 & 1.11 \\
5.0      &  0.0434 & -0.0101 & 1.09 & 1.05 & 1.03 \\
2.5      &  0.0707 &  0.0374 & 1.14 & 1.05 & 1.13 \\
1.8      &  0.0675 &  0.0536 & 1.13 & 1.02 & 1.13 \\ \hline
\end{tabular}
\begin{tabular}{|c|c|c|c|c|c|}
\hline
\multicolumn{6}{|l|}{$Z_{Q}$} \\ \hline
$m_{Q}$ & \makebox[15mm]{$C-A$}            &
          \makebox[15mm]{$\widetilde{C-A}$}    &
          \makebox[15mm]{$Z_{Q}^{pert}$} &
          \makebox[15mm]{$Z_{Q}^{mf}$}   &
          \makebox[15mm]{$Z_{Q}^{tad-imp}$} \\ \hline
$\infty$ &  0.0383 &  0.0383 & 1.075 & 1.000 & 1.075 \\
5.0      &  0.0673 &  0.0173 & 1.132 & 1.077 & 1.114 \\
2.5      &  0.1003 &  0.0003 & 1.197 & 1.161 & 1.162 \\
1.8      &  0.1268 & -0.0121 & 1.249 & 1.235 & 1.206 \\ \hline
\end{tabular}
\end{center}
\caption{Renormalization parameters for non-relativistic QCD.
         The $n$=1 action is used at $m_{Q}$=$\infty$ and 5.0
         while the $n$=2 action is used at 2.5 and 1.8.
         As the  coupling constant we use
         $g_{V}^{2}(\pi/a)$=1.96 at $\beta$=6.0.
         }
\label{ren_params}
\end{table}
\begin{table}
\begin{center}
\begin{tabular}{|c|c|c|c|c|}
\hline
\multicolumn{5}{|l|}{$Z_{A}$} \\ \hline
$m_{Q}$ & \makebox[15mm]{$V^{latt}$}  &
          \makebox[15mm]{$\tilde{D}$} &
          \makebox[15mm]{$1/Z_{A}^{latt}$} &
          \makebox[15mm]{$Z_{A}$}  \\ \hline
$\infty$ & 0.1070 & 0.1346 & 0.638 & $\infty$ \\
5.0      & 0.0972 & 0.1143 & 0.654 & 0.721 \\
2.5      & 0.0920 & 0.1006 & 0.655 & 0.692 \\
1.8      & 0.0897 & 0.0921 & 0.650 & 0.671 \\ \hline
\end{tabular}
\end{center}
\caption{Renormalization factors for the heavy-light
         axial vector current.
         As a coupling constant we use 
         $g_{V}^{2}(2.18/a)$=2.22\protect\cite{Hernandez_Hill}
         at $\beta$=6.0.
         }
\label{Z}
\end{table}
\begin{table}
\begin{center}
\begin{tabular}{|c|c|c|c|c|}
\hline
&
\makebox[10mm]{$m_{Q},n$}                &
          \makebox[30mm]{$(f_{P} \sqrt{m_{P}}/Z_{A})^{static}$} &
          \makebox[30mm]{$c$} &
          \makebox[30mm]{$f_{B} \sqrt{m_{B}}/Z_{A}$} \\ \hline
(a) &
     \begin{tabular}{c}
     10.0, 7.0 ($n$=1) \\
     and \\
     5.0, 4.0, 3.0, \\
     2.5 ($n$=2)
     \end{tabular}
    & 0.292(35) & 1.04(35) & 0.124(44)\\ \hline
(b) &
     \begin{tabular}{c}
     10.0, 7.0, 5.0,\\
     4.0 ($n$=1)
     and \\
     3.0, 2.5 ($n$=2)
     \end{tabular}
    & 0.284(31) & 0.61(21) & 0.187(19) \\ \hline
(c) &
     \begin{tabular}{c}
     10.0, 7.0, 5.0,\\
     4.0 ($n$=1)
     \end{tabular}
    & 0.307(44) & 1.03(44) & 0.132(52)\\ \hline
(d) &
     \begin{tabular}{c}
     5.0, 4.0, 3.0, \\
     2.5 ($n$=2)
     \end{tabular}
    & 0.262(22) & 0.64(21) & 0.168(22) \\ \hline
\end{tabular}
\end{center}
\caption{Results of fitting with the form
         (\protect\ref{fit:equation}) in the text and the value of 
         $f_{P} \protect\sqrt{m_{P}}/Z_{A}$ at $m_Q=1.8$ (last column).
         (a), (b), (c) and (d) label the selection of
         data points.
         }
\label{m-inv-fit}
\end{table}
\begin{table}
\begin{center}
\begin{tabular}{|c|c|c|c|}
\hline
Group                & Lattice & Smearing &
                        $(f_{P} \sqrt{m_{P}}/Z_{A})^{static}$ \\
\hline
APE\cite{APE_static} & $\mbox{18}^{3}\times\mbox{32}$
                     & cube, $n$=5    & 0.368(14) \\
                   & &       $n$=7    & 0.311(14) \\ \hline
Bernard et al.\cite{BLS_static} & $\mbox{24}^{3}\times\mbox{39}$
                     & cube, $n$=5    & 0.369(18) \\
                   & &       $n$=7    & 0.299(10) \\
                   & &       $n$=9    & 0.241(12) \\ \hline
PSI-Wuppertal\cite{PSW_static} & $\mbox{12}^{3}\times\mbox{36}$
                     & exponential, gaussian   & 0.327(14) \\ \hline
UKQCD\cite{UKQCD_heavy-light} & $\mbox{16}^{3}\times\mbox{48}$
                     & Jacobi &                  0.298(10) \\ \hline
this work            & $\mbox{16}^{3}\times\mbox{48}$
                     & independent (see text.) & 0.292(35) \\ \hline
\end{tabular}
\end{center}
\caption{Comparison of results for 
         $(f_{P} \protect\sqrt{m_{P}}/Z_{A})^{static}$
         obtained by several groups at $\beta$=6.0.
         }
\label{comparison:data}
\end{table}

\newpage
\begin{figure}[h]
\hspace*{-0.5cm}
\epsfxsize=16cm
\epsffile{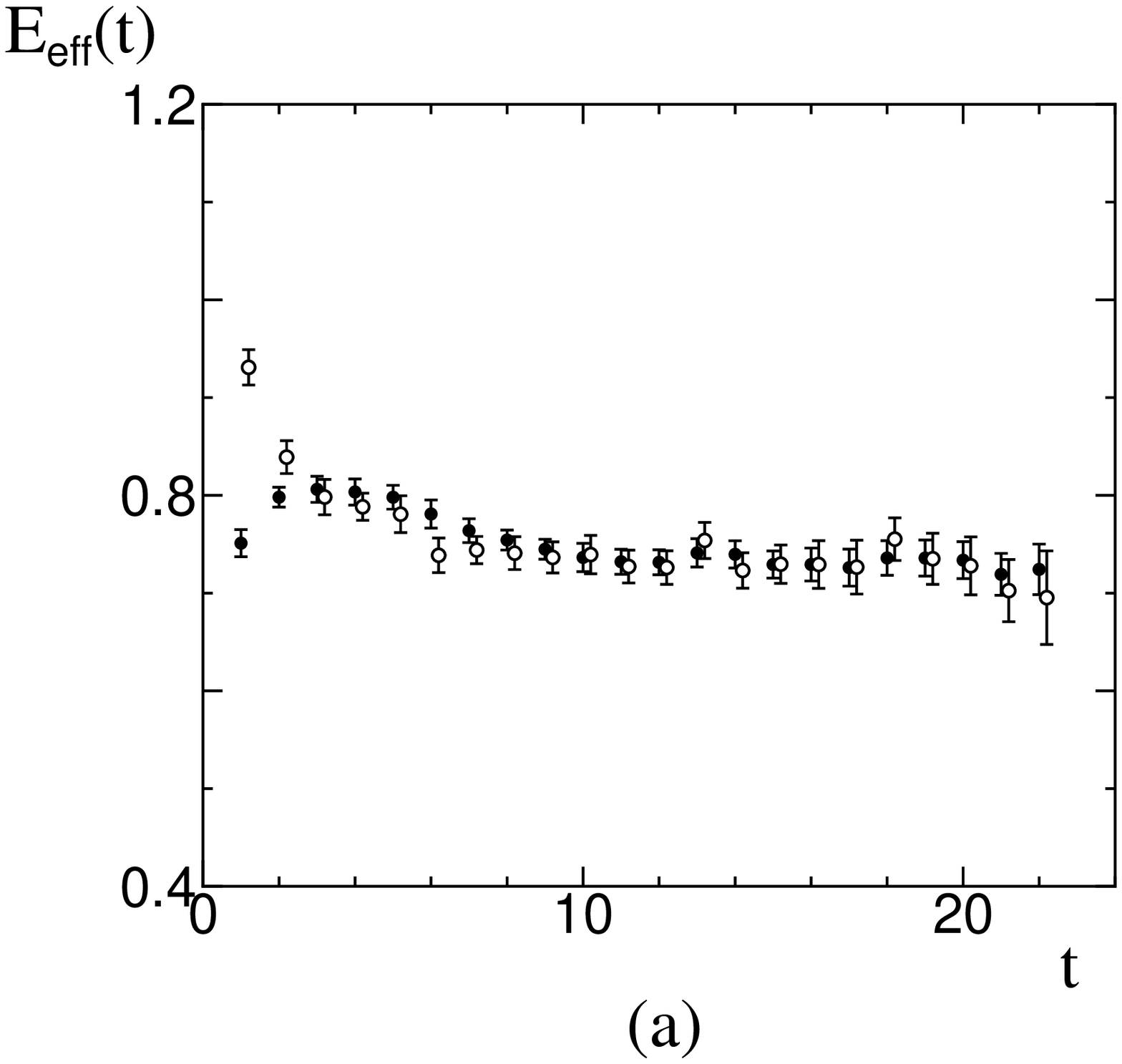}
\end{figure}

\begin{figure}[h]
\hspace*{-0.5cm}
\epsfxsize=16cm
\epsffile{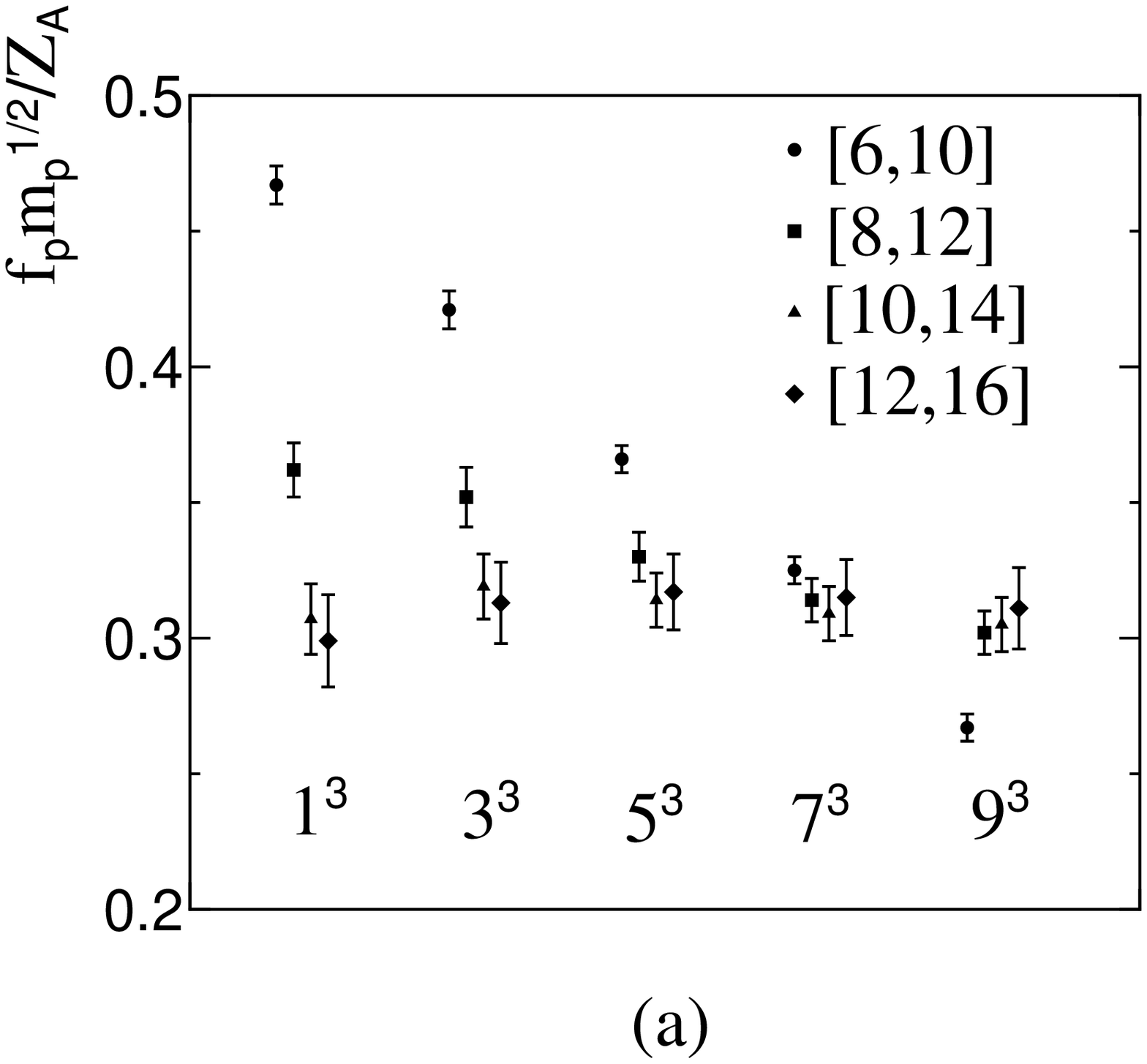}
\end{figure}


\end{document}